\documentclass[letter]{spie}  
\usepackage[]{graphicx}
\usepackage[dvips]{hyperref}
\def\txt#1{{\mbox{\scriptsize #1}}}

\title{Neodymium Photoluminescence in Whispering Gallery Modes of Toroidal Microcavities}

\author{Fedja Orucevic, Jean Hare and Val\'erie Lef\`evre-Seguin
\skiplinehalf 
Laboratoire Kastler Brossel -- \'Ecole normale supérieure; UPMC; CNRS\\
 24 rue Lhomond, F-75231 Paris Cedex 05, France}

\authorinfo{Corresponding author \url{Fedja.Orucevic@lkb.ens.fr}\\
{\scriptsize This paper was published in "Optoelectronic Devices: Physics, Fabrication, and Application III", \emph{Proceedings of SPIE}, Vol. 6368 (2006)  and is made available as an electronic reprint  with permission of SPIE. One print or electronic copy may be made for personal use only. Systematic or multiple reproduction, distribution to multiple locations via electronic or other means, duplication of any material in this paper for a fee or for commercial purposes, or modification of the content of the paper are prohibited.}}

\begin{document}
  \maketitle

\begin{abstract}
  Copyright 2006 Society of Photo-Optical Instrumentation Engineers.
\end{abstract}



\section{INTRODUCTION}
\label{sect:intro}  

The pioneering studies on the interaction between an electromagnetic field and
its surrounding environment were conducted by Purcell in 1946
\cite{QC_PURCELL46} and Casimir in 1948 \cite{QC_CASIMIRPOLDER48}. Since then
this field of physics has drawn growing interest in both fundamental physics
and applications. In the past three decades the use of optical microcavities in
the domains such as cavity quantum electrodynamics (CQED) \cite{ENS_HOUCHES90},
non-linear optics, quantum states engineering, etc. led to numerous experiments
that enhanced the comprehension of quantum physics and its fundamental concepts
such as decoherence \cite{QM_ZUREKDECOH81,ENS_REVERSIBLEDECOHERENCE},
entanglement \cite{ENS_CATPHIL,QO_KNIGHTCAT92}, quantum non demolition
measurement (QND)\cite{ENS_QNDPROPOSAL}, etc.

The apparition of semiconductor microcavities allowed to reproduce
those results in the optical domain as required for applied devices
\cite{YoshieScherer04}. These cavities, generally obtained from MBE
grown III-V heterostructures by suitable etching, are mostly of
three kinds, named after their shape as micropillars (Fabry-Perot
cavities with Bragg reflectors with lateral confinement Total
Internal Reflections), photonic crystals (achieving light
confinement by 2D or 3D Bragg reflection), and microdisks, using
whispering-gallery modes (described below). These cavities generally
achieve very good spatial confinement, with mode volumes close to
the fundamental limit of $(\lambda/N)^{3}$ (where $\lambda$ is the
vacuum wavelength and $N$ the refractive index), and their quality
factors have been increased progressively from a few hundred to more
than tens of thousands, still limited by surface roughness resulting
from the etching process. Semiconductor technologies used in
fabrication of these cavities make them ideal candidates for
integrated optoelectronic devices.

Alternatively, dielectric optical microcavities, such as fused
silica microcavities have also been studied as they achieve very
high quality factors (up to $10^{10}$). They can have different
shapes (spherical, cylindrical, toroidal...) but the presence of
revolution symmetry allow them to support the so called ``Whispering
Gallery Modes'' (WGM) resulting from light guiding below the curved
surface by successive Total Internal Reflections (TIR) at grazing
incidence. The very high quality factors  of WGM, as high as
$10^{10}$, as measured in Ref.
\citenum{BraginskyGorodetsky89},\citenum{CollotLefevre93},
\citenum{GorodetskySavchenkov96} and \citenum{VernooyIlchenko98},
result from the intrinsic high transparency of the silica and very
small surface roughness. Combined with small modal volumes (few
hundreds of $(\lambda/N)^{3}$), they lead to unequalled $Q/V$ ratio
and make these cavities perfectly suited for low scale, low
threshold lasers and non-linear optical devices.

The coupling between dielectric cavities and different kind of emitters has
been extensively studied and reported in the previous works
\cite{ChangCampillo96, MatskoSavchenkov05}. Using the electrical dipole
approximation, in the weak coupling regime the rate of spontaneous emission is
given by Fermi golden rule as $\Gamma \propto \rho(\omega) \left|
\textbf{d}\cdot\textbf{E}(\omega)\right|^{2}$, where $\rho(\omega)$ stands for
electromagnetic mode density, $\textbf{d}$ for dipole strength and
$\textbf{E}(\omega)$ for electrical field. Therefore, the most efficient
coupling can be achieved with great dipole emitters placed in the region of
maximal electric field. Alkali atoms (namely rubidium or cesium atoms) and
semiconductor heterostructures (InAs/GaAs self-assembled quantum dots), have
large dipoles (1-10 atom units) but cannot be embedded inside the silica matrix
where the WGM field reaches its maximum. However, they can be coupled to WGM
using the evanescent wave \cite{HareSteiner05}.

In this work we used the microtoroidal cavities, first introduced by Vahala and
co-workers \cite{ArmaniKippenberg03}. They are obtained by the laser fusion of
silica microdiscs and combine the advantages of easier microdisks integration
on silicon substrates and higher $Q/V$ ratios of dielectric fused silica
cavities. We used the neodymium ions because they can be easily embedded in the
silica matrix by using ionic implantation technique. Despite their small dipole
($d\sim 10^{-2}\;e a_{0}$) and the large phonon induced homogenous broadening
(at room temperature), the coupling of a large number of Nd ions with the
maximum electric field, allows to easily observe cavity effects.

\section{FABRICATION OF MICROTOROIDS}

\begin{figure}
\begin{center}
\includegraphics[width=100mm]{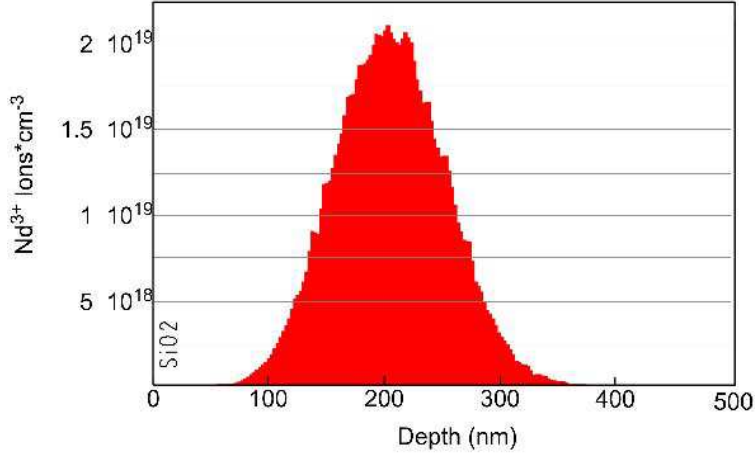}
\end{center}
\caption[NdImp]{\label{NdImp}
th distribution of implanted $\textrm{Nd}^{3+}$ ions as calculated with SRIM
tware. The maximum density of $\textrm{Nd}^{3+}$ ions occurs at the depth of
 nm and FWHF is about 110~nm.}
\end{figure}

Fabrication of microtoroids is a two-step process involving
successively silicon microelectronic technology and
$\textrm{CO}_{2}$ laser fusion of silica. In the first step a 800 nm
thick layer of silica is produced by thermal oxydation of a silicon
substrate. Then silica is bombarded with
600~keV-$\textrm{Nd}^{3+}$-ions with fluency of $2.5\cdot10^{14}$
$\textrm{ions}\cdot\textrm{cm}^{-2}$. The distribution of
$\textrm{Nd}^{3+}$ ions in matrix is calculated using SRIM software
\cite{BiersackHaggmark80,Ziegler85} and yields quasi-gaussian
concentration distribution shape with the peak of $2\cdot10^{19}$
$\textrm{ions}\cdot\textrm{cm}^{-3}$ at the depth of 200 nm
(Fig.\ref{NdImp}). The limited power of the accelerator did not
allow to reach the depth of 400 nm (i.e. the center of the silica
layer) where the maximum WMG field is expected. However, the large
width of neodymium ion distribution still permitted very good
coupling with WGMs (see subsequent paragraphs). After implantation,
circular photoresist pads produced by optical or e-beam lithography
are transferred to the silica layer by wet etching in buffered HF.
The edge of the resulting silica disk is then isolated from the
substrate by $\textrm{SF}_{6}/\textrm{O}_{2}$ selective isotropic
etching.

\begin{figure}
\begin{center}
\includegraphics[width=150mm]{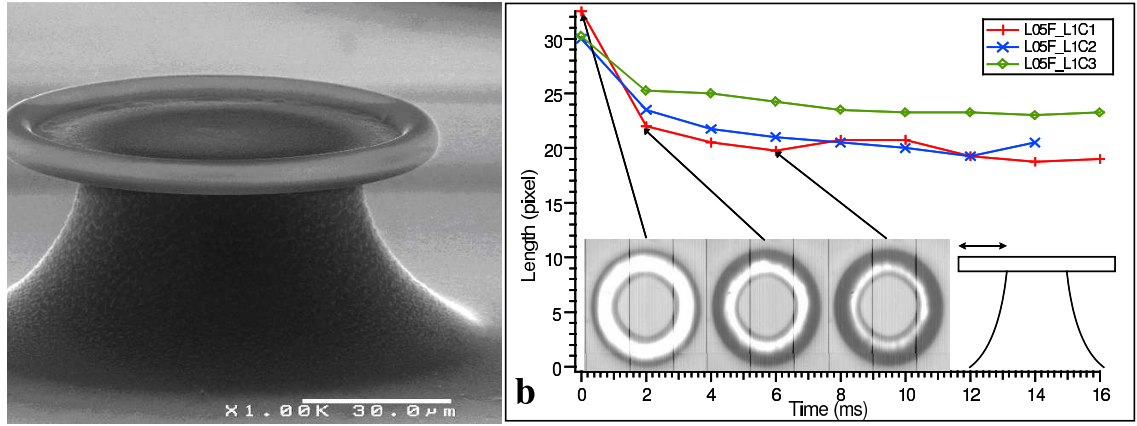}
\end{center}
\vskip 5mm
\caption[Dynami]{ \label{Dynami} \textbf{a}~---~SEM image of a pure silica 90 $\mu \textrm{m}$ diameter microtoroid. \textbf{b}~---~The curves show the undercut length of a
several microdiscs as a function of irradiation time. The undercut is defined
as depicted by the double arrow on the schema at the bottom right corner of the
figure. In the inlet are shown three images of a given microdisc at three
different times of fusion process (0 ms, 2 ms and 6 ms).}
\end{figure}

In the second step we use $\textrm{CO}_{2}$ laser to irradiate the
microdisks. As silicon pillar acts as a very efficient thermal sink
\footnote{The temperature varies less than 100 K between top side,
in contact with silica, and bottom side, in contact with silicon
substrate acting as thermostat.} in the center of the microdisk the
resulting radial temperature gradient leads to fusion of silica only
on the microdisk's edges. Surface tension acting on the molten
material gives the final toroidal shape of these microcavities
(Fig.\ref{Dynami}-a). Using a fast camera (1000 fps) we observed the
dynamics of the microtoroids' formation. The fusion process leads to
the final shape of microtoroid in less than 10 ms
(Fig.\ref{Dynami}-b). Using the cylindrical symmetry, we easily
modeled heat transfer process in microdiscs. Dimensions of the
microdiscs that we used (typically, diameter $d=100\,\mu \textrm{m}$
and disc thickness $t=800\,\textrm{nm}$) lead to very short
characteristic times of heat diffusion through silica
($\tau_{SiO_{2}}=8\,\mu\textrm{s}$) and silicon
($\tau_{Si}=25\,\mu\textrm{s}$) that are much smaller than the
microtoroids' formation time. Therefore the Fourier time-dependant
heat transfer equation is recast in a Poisson-like steady-regime
equation that we resolved using a standard over-relaxation method.

The formation of a microtoroid is an auto-regulated process: the molten silica
reflow toward center until the temperature of the microtoroid's edge reach the
fusion point. Thus, the $\textrm{CO}_{2}$ laser power and the geometry of
microdiscs solely determine the temperature gradient and hence the final shape
of a microtoroid, as far as the duration of the laser shot is greater than the
time required to form the microtoroid ($\simeq 10\,\textrm{ms}$).

The pillars of $\textrm{Nd}^{3+}$ implanted microdiscs  had not the perfect
cylindrical symmetry, but presented a square-like shape on the top side in
contact with the silica microdiscs. That could be the result of the
$\textrm{Si}/\textrm{SiO}_{2}$ contact interface as the pillars are etched
preferentially over the two Si crystal axis. Thanks to the basics properties of
heat diffusion, the temperature distribution which is square-shaped on the
pillars becomes more circular further from the center. We managed to stop the
formation of microtoroid before it approached the pillars too much by
controlling the $\textrm{CO}_{2}$-laser pulse via a computer generated TTL
gate. The succession of short manually triggered 1~ms-pulses allowed to control
distance between toroid edge and the pillar.

\section{EXPERIMENTAL PROCEDURE}

\begin{figure}[th]
\begin{center}
\includegraphics[width=130mm]{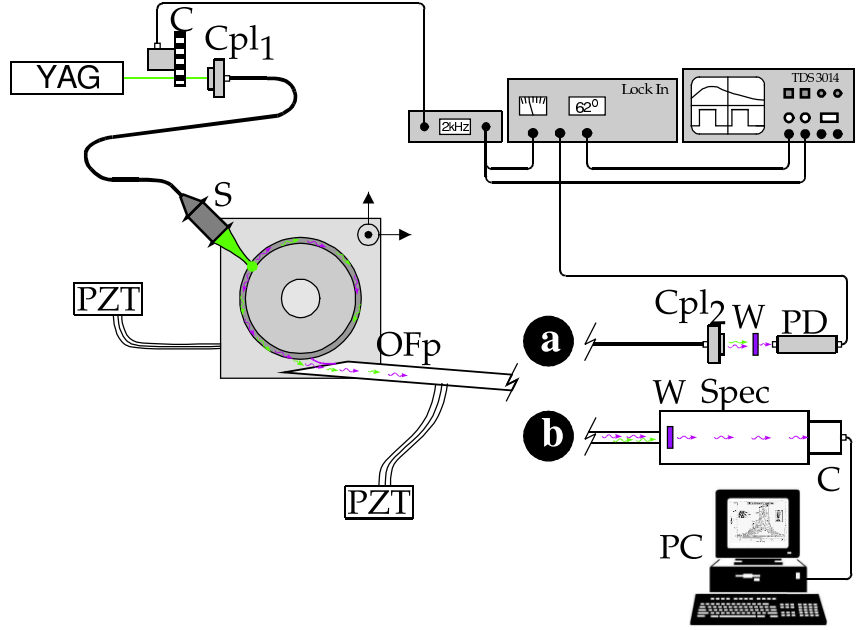}
\end{center}
\caption[ExpSet]{ \label{ExpSet} Sketch of experimental setup. The pump laser is a doubled YAG
($\lambda=532\,\textrm{nm}$), coupled ($\textrm{Cpl}_{1}$) into the singlemode
fiber. A system (S) of two lenses is used to focus the laser beam to a spot of
 $6\,\mu\textrm{m}$ in diameter. PL signal from the $\textrm{Nd}^{3+}$
embedded ions is collected with the polished fibre
($\textrm{OF}_{\textrm{p}}$).
  \textbf{Path~a}~--~PL is collimated by coupler ($\textrm{Cpl}_{2}$) to
the photodiode (PD) through a $850\,\textrm{nm}$ cut-off filter (W)
to remove the residual pump light. \textbf{Path~b}~--~After
optimisation, the PL is sent through the spectrograph's slits. A
$850\,\textrm{nm}$ cut-off filter (W) is placed inside the
spectrograph (Spec) and the signal is recorded with the
spectroscopic camera (C).}
\end{figure}

We pumped neodymium ions by a tightly focused laser on edge of microtoroid
($\lambda = 532\,\textrm{nm}$, pump waist $\simeq 3\,\mu\textrm{m}$,
$P=9\,\textrm{mW}$). This enhances the detection contrast between  WGM confined
near the microtoroid's edge and leaky modes located in the whole volume of
microtoroid.

We first performed the collection of light scattered from a microtoroid using a
cleaved multimode optical fiber placed about $100\,\mu\textrm{m}$ apart from
it. The spectrum of the scattered light does not exhibit resonances but matches
the PL spectrum of neodymium ions embedded in bulk silica. Indeed, most of
$\textrm{Nd}^{3+}$ ions  are not coupled to the WGMs but rather  emit in the
leaky modes of the cavity and the large numerical aperture fiber ($NA\simeq
0.22$) placed in the far field collects light originating from the large
portion of the microtoroid. In a previous work \cite{VerbertMazen05}, a
microscope objective allowed to selectively detect light coming from the edge
of erbium doped microtoroids, thus leading to observe the WGM emission
spectrum.

Here, we chose to perform this selective detection by a near field method. This
was done by evanescent wave coupling between the microtoroid and an angle
polished multi-mode fiber \cite{IlchenkoYao99}. The angle of polishing $\Phi$
is chosen to fulfill the phase-matching requirement for WGMs:
\begin{equation}
\Phi=\arcsin \left( \frac{N_\txt{eff}}{N_\txt{f}} \right), \label{PhiPol}
\end{equation}
\noindent where $N_\txt{f}$ is the  fiber core index and $N_\txt{eff}$ the
effective index of the microtoroid's WGM, that can be estimated from the
semiclassical equation for a sphere:
\begin{equation}
Nx=\ell + \frac{1}{2} + \left(\frac{\ell +1}{2} \right)^{1/3} \alpha_{n} +
\cdots \label{PosRes}
\end{equation}
\noindent Here $N$ is the refractive index of silica, $x={2\pi a}/{\lambda}$
the size parameter, where $a$ is the radius of the sphere, $\ell$ is the
quantum angular number and $\alpha_{n}$ are the successive zeros of $Ai(-z)$
(Airy function). For large size parameters ($\ell\gg1$) and low order WGMs ($n$
close to 1), the internal caustic is close to the surface and we can substitute
$\ell \simeq Nx$ in the equation (\ref{PosRes}) to obtain:
\begin{equation}
N_\txt{eff}\equiv \frac{\ell}{x} \approx N \left[ 1-
\frac{2^{-1/3}}{(Nx)^{2/3}}\;\alpha_{n} \right]. \label{EffInd}
\end{equation}
\noindent For a microtoroid's diameter of $50\,\mu\textrm{m}$ and
for the $n=1$ radial order WGM the angle is $\Phi\simeq 18^{\circ}$.
As the microtoroids hold on pillars of $20-30\,\mu\textrm{m}$ in
height from silicon substrate, one needs to use a fiber with
suitable core and cladding diameters, i.e.
$(d_\txt{cl}-d_\txt{co})/2<h_\txt{p}$. Here $d_\txt{cl}$ and
$d_\txt{co}$ are the diameters of, respectively, the optical
cladding and the fiber core, while $h_\txt{p}$ is the height of a
pillar. In our experiment we used a fiber with
$d_\txt{co}=100\,\mu\textrm{m}$, $d_\txt{cl}=110\,\mu\textrm{m}$ and
$\textrm{NA}=0.22$ (Newport F--MCB--T).

Fig. \ref{ExpSet}-a shows the experimental setup used to optimise
the PL signal. Two sheets (W) of $850\,\textrm{nm}$ cut-off filter
(Wratten 87C) remove residual pump light before the signal reaches a
Si-photodiode (PD). To extract the PL signal from the noise, the
pump beam was chopped (C) at a frequency of about 2.5 kHz and a
lock-in amplifier was used. Finally, the signal is monitored on the
oscilloscope and optimised using PZT to control the position of
laser spot and the gap between the microtoroid and the fiber. Then,
the signal analysed with a grating spectrograph (Acton SpectraPro
300i, 1200 grooves/mm) is accumulated for 10~min on a cooled
spectroscopic camera (Princeton Instruments SpectruMM 120) as shown
in the Fig. \ref{ExpSet}-b.

\section{RESULTS AND DISCUSSION}
\begin{figure}[b]
\begin{center}
\includegraphics[width=130mm]{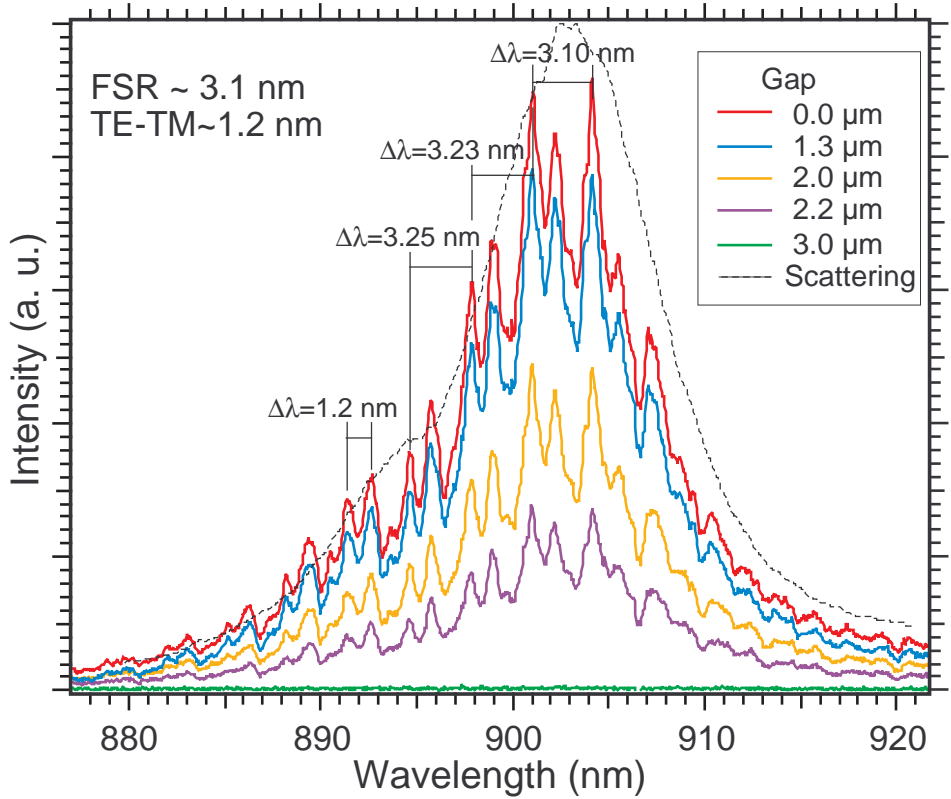}
\end{center}
\caption[WGM]{ \label{WGM} PL spectra from a $\textrm{Nd}^{3+}$ doped
$55\,\mu\textrm{m}$ diameter microtoroid. The dotted curve is the PL
spectrum of scattered light from the $\textrm{Nd}^{3+}$ ions. Other
curves are the PL spectra obtained by coupling the evanescent wave.
Two frequency intervals of $3.2\,\textrm{nm}$ and $1.2\,\textrm{nm}$
correspond, respectively, to the cavity FSR and TE-TM frequency
splitting.}
\end{figure}

The expected Free Spectral Range (FSR) of WGM in the microtoroids (and more
generally, any other cylindrical symmetry cavity) is given by
\begin{equation}
\Delta \lambda_\txt{FSR} =\frac{\lambda^2}{2\pi\,a\,N_\txt{eff}},
\label{FsrExp}
\end{equation}
\noindent where $a$ is the microtoroid's great radius, $N_\txt{eff}$
is the effective index as obtained in eq. \ref{EffInd} and $\lambda$
is the vacuum wavelength. Fig.~\ref{WGM} shows PL spectra obtained
from a microtoroid of about $55\,\mu\textrm{m}$ in diameter
(measured with standard optical microscope). The different curves
correspond to different values of the gap between the fiber and the
microtoroid. We identify a FSR of about $3.2\,\textrm{nm}$, as
expected from equation \ref{FsrExp} for $55\,\mu\textrm{m}$ diameter
microtoroids. The frequency interval of about $1.2\,\textrm{nm}$
corresponds to the TE-TM polarisation splitting. The signal strength
is maximum when the fiber is in contact with microtoroid and
disappears at $3\,\mu\textrm{m}$ gap.

The ultimate spectrograph resolution of $0.1\,\textrm{nm}$ restrains us from
measuring quality factors higher than a few thousands, while we expect values
in the range of $10^{8}-10^{10}$ \cite{ArmaniKippenberg03}. Furthermore, as the
fiber core diameter ($80\,\mu\textrm{m}$) is very large compared to the
thickness of the microtoroid (a few micrometers), it is very likely that
several WGMs are coupled in the fiber at once. Those factors contribute to the
broadening of the peaks observed in the spectrum.

\section{PROSPECTS}

The peak of neodymium absorption is at 810 nm, so the pump efficiency can be
improved using a laser at this wavelength rather than 532 nm. The development
of tapered fibre couplers, that we have just achieved, will even allow to
inject the pump laser in resonance with a WGM and therefore enhance the
luminescence of neodymium ions in the WGMs.

Our next aim is to demonstrate low threshold Nd-microlaser as an integrated
version of an earlier reported work on microspheres
\cite{SandoghdarTreussart96} and the observation of Silicon Rich Oxides (SrO)
microtoroids' PL spectrum.

\section*{ACKNOWLEDGMENTS}

The authors would like to warmly acknowledge J\'er\'emy Verbert,
Emmanuel Hadji and Jean-Michel G\'erard from SP2M - CEA - Grenoble
for very rich and fruitful collaboration and for providing
microdiscs samples.

This work has been supported by the ``Programme National Nanosciences''
(ACI~Microtores).

\bibliographystyle{spiebib}   

\end{document}